\documentclass[5p]{elsarticle}

\usepackage[utf8]{inputenc}
\usepackage{maltese}
\usepackage{hyperref}
\usepackage[all]{hypcap}

\journal{Journal of Archaeological Science: Reports}

\begin{document}

\begin{frontmatter}

\title{The Frequency Spectrum and Geometry of the {\mH}al Saflieni Hypogeum Appear Tuned}

\author[address1]{Kristina Wolfe}
\author[address2]{Douglas Swanson}
\author[address1]{Rupert Till}

\address[address1]{University of Huddersfield}
\address[address2]{Independent Researcher}

\begin{abstract}
The {\mH}al Saflieni Hypogeum is a unique subterranean Maltese Neolithic sanctuary with a well-documented history of interest in its acoustics \cite{Till, Debertolis, Reznikoff2014, Devereux2009, Skeates, Mifsud, Devereux1996, Evans, Zammit, Griffiths}.  Previous studies have noted its unusual strongly-defined frequency spectrum \cite{Debertolis}, but it is unknown if this was coincidental.  In this paper, we present evidence that the Hypogeum’s creators shaped the site’s geometry to create or amplify its frequency spectrum, or another property closely correlated with the spectrum.  Specifically, we show that the observed spectrum required jointly fine-tuning the dimensions of multiple non-contiguous cave walls across multiple independent chambers, to a degree that seems unlikely to be coincidental.  We also note that the peak frequencies are evenly spaced and resemble a whole-tone scale in music, which is also unlikely to be coincidental and suggests the spectrum itself might have held some cultural significance.  Taken together, it suggests acoustic or spectral properties may have played a motivational or cultural role for the site’s Neolithic creators.  This work identifies one of the earliest known examples of a manmade structure with a significant musical element to its interior architecture.
\end{abstract}


\end{frontmatter}

\section{Introduction}

The {\mH}al Saflieni Hypogeum is a subterranean Neolithic structure and UNESCO World Heritage Site in Paola, Malta \cite{Till, Debertolis, Reznikoff2014, Devereux2009, Skeates, Mifsud, Devereux1996, Evans, Zammit, Griffiths, HeritageMalta, Google}.  It is among the oldest and best-preserved manmade structures with an intact interior.  Hand-carved with stone and bone tools, it is believed to have been a sanctuary and then an ossuary \cite{Mifsud, Evans, Zammit}.  Previous studies have noted its unique acoustic properties, such as low-frequency amplification, long reverberation times, and well-defined peak frequencies \cite{Till, Debertolis, Reznikoff2014}, but it is unknown whether these properties were coincidental.

In this paper, we present the first comprehensive measurements of the frequency spectrum of the Hypogeum middle level.  We identify strong well-defined peak frequencies common to multiple non-contiguous chambers.  We show that these peak frequencies can be reproduced by a simulation of the 3D wave equation incorporating only the site’s geometry, suggesting they represent room modes.  We further show that prominent peaks can be modeled analytically as the modal frequencies of individual chambers treated as rectangular boxes, even though the chambers are in reality highly non-rectangular.  We use this simplified “box model” to estimate the distances that individual chamber walls could be perturbed before prominent peaks would shift by human-noticeable amounts.  We validate these estimates by re-simulating the 3D wave equation in a perturbed version of the full non-rectangular geometry.

Our results suggest the aggregate spectrum across the Hypogeum middle level required jointly fine-tuning the dimensions of multiple non-contiguous chamber walls to within 10-25 cm.  This seems unlikely to have come about by coincidence, and suggests that the Hypogeum’s Neolithic creators shaped the site’s geometry to create or amplify its frequency spectrum, or another property closely correlated with the spectrum.

\section{Methodology}

The Hypogeum consists of three levels:  upper (3600-3300 BC), middle (3300-3000 BC), and lower (3150-2500 BC) \cite{Trump}.  The field work in this paper was undertaken in the inner chambers of the middle level.  A diagram of the middle level is shown in Figure \ref{figure1a}.  A well-known early 20th century photograph of Chamber 26 from \cite{Griffiths} is shown in Figure \ref{figure1b} as an example of the hand-carved ornamentation and smooth Lower Globigerina limestone surfaces common to middle level chambers.  Additional photographs and descriptions of the site and its material composition may be found in \cite{Mifsud, Evans, Zammit, Griffiths}.  Virtual video tours to help visualize the site layout may also be found in \cite{HeritageMalta, Google}.

\renewcommand{\thefigure}{1(a)}
\begin{figure*}[h!]
\centering
\includegraphics[width=0.8\textwidth]{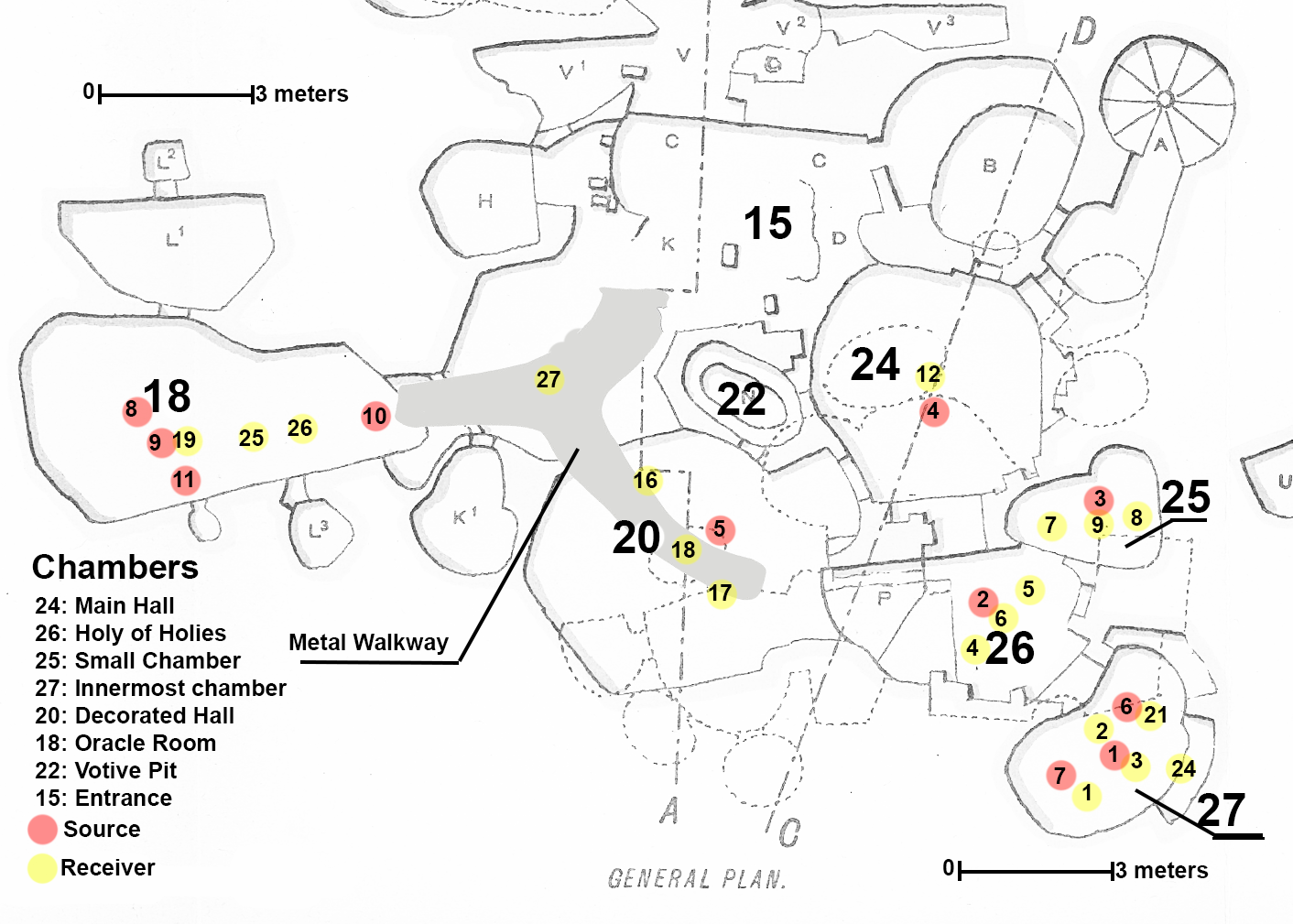}
\caption{Diagram of the Hypogeum middle level from \cite{Zammit} with approximate positions of experimental sound sources and receivers.  Chamber numbering is from \cite{Evans}.}
\label{figure1a}
\end{figure*}

\renewcommand{\thefigure}{1(b)}
\begin{figure*}[h!]
\centering
\includegraphics[width=0.8\textwidth]{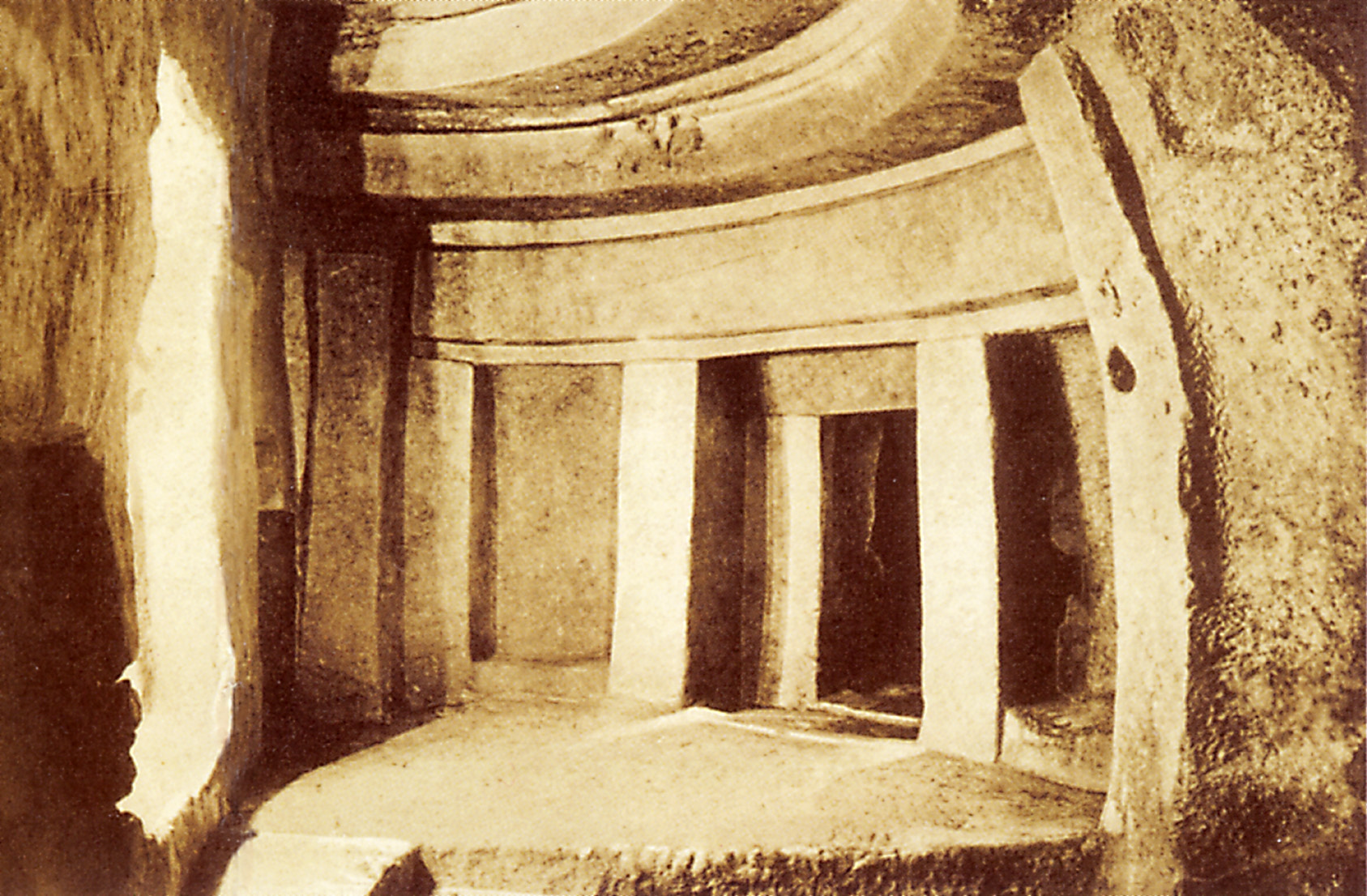}
\caption{A well-known early 20th century photograph of Chamber 26 from \cite{Griffiths}.  Additional photographs and descriptions of the site and its material composition may be found in \cite{Mifsud, Evans, Zammit, Griffiths}.  Virtual video tours to help visualize the site layout may also be found in \cite{HeritageMalta, Google}.}
\label{figure1b}
\end{figure*}

We measured the frequency spectrum using standard acoustic methods.  Impulse responses were generated using the sine sweep measurement method \cite{Muller} at several source (loudspeaker) and receiver (microphone) positions throughout the middle level.  Measurements were taken following guidelines for the acoustic analysis of rooms and performance spaces outlined in ISO-3382-1 \cite{ISO3382, Fazenda}.  This was deemed most relevant given the archaeological record.  Although ISO-3382-1 is primarily focused on analysis of music performance venues, its acoustic quantities can also characterize other forms of listener experience \cite{Watson, Pentcheva, Rainio}.  Ambient environmental conditions during the field work including temperature and humidity were provided by Heritage Malta in order to estimate the speed of sound for use in simulations and modeling.  Source and receiver positions were chosen based on archaeological context, recommendations from Heritage Malta archaeologists, and results from previous field work \cite{Till}.  For each source and receiver position, a 45-second 48-kHz 32-bit exponential sine sweep was generated within the range of 20Hz to 22.5kHz with a tail length of 17 seconds.  The sound source was a Bang \& Olufsen Beolit 12 loudspeaker \cite{Beolit}.  Sweeps were recorded using two omnidirectional microphones (Neumann KM-D digital microphone with Neumann KK-131 diaphragm) and an ambisonic sound field microphone (Sennheiser Ambeo-VR), through two MOTU Ultralight sound cards.  Impulse responses were recorded and processed using the HAART \cite{Johnson} library in Max \cite{Max}.

Frequency spectra were extracted using the Fast Fourier Transform over the entire 17-second impulse response, and smoothed using a Savitzky-Golay filter to improve visual display \cite{Smith, Savitzky}.  Mean frequency spectra were calculated for each chamber by averaging all impulse responses from source-receiver pairs in that chamber, and then extracting the frequency spectrum of the mean impulse response.  A Hypogeum-wide mean spectrum was calculated by averaging the mean impulse responses from chambers 18, 20, 24, 25, 26, and 27, so that all chambers contributed equally regardless of their numbers of sources and receivers, and then extracting the frequency spectrum of this Hypogeum-wide mean impulse response.  This Hypogeum-wide mean spectrum provided a quantitative representation of the acoustic behavior of the site.

To reproduce the observed frequency spectra and explore the effects of perturbing the site’s geometry, we simulated the 3D wave equation.  A high-resolution digital geometric model of the site provided by Heritage Malta was simplified and remeshed using MeshLab \cite{MeshLab}.  Simulations were implemented in C++ using libigl \cite{libigl} following the approach of Bilbao and Hamilton \cite{Bilbao, Hamilton}.  Simulations modeled rigid wall boundary conditions and Kronecker delta initial conditions and did not incorporate any material properties, absorption coefficients, or viscothermal relaxation, as they were not found to be necessary to reproduce the experimental spectra.  Simulated impulse responses were generated and simulated frequency spectra extracted in the same manner as the experimental frequency spectra.

To estimate the perturbation size (the change in chamber wall dimensions) necessary to cause individual chamber peak frequencies to deviate noticeably from the Hypogeum-wide mean spectrum, we modeled the chambers analytically as rectangular boxes.  This “box model” was chosen for its simple and analytically-solvable room modal frequencies.  Approximate rectangular dimensions (length, width, height) for each chamber were estimated using MeshLab \cite{MeshLab} and resulting room modal frequencies were calculated using the standard equation: \[ f = \frac{c}{2} \sqrt{ \frac{n_x^2}{L_x^2} + \frac{n_y^2}{L_y^2} + \frac{n_z^2}{L_z^2}} \] where $n_x$, $n_y$, and $n_z$ are modal indices, $L_x$, $L_y$, and $L_z$ are box dimensions, and $c$ is the speed of sound \cite{Bilbao}.  Differentiating with respect to $L_i$ leads to a similar equation governing the response of a room modal frequency to a perturbation in a box dimension: \[ \Delta f = \frac{-c^2 n_i^2}{4 f L_i^3} \Delta L_i \]

We calculated the perturbation size $\Delta L_i$ needed to produce a modal frequency deviation $\Delta f$ of 3 Hz, the approximate just-noticeable difference for modern humans for frequencies below 500 Hz \cite{Kollmeier}.\footnote{We note that 3 Hz is likely a conservative estimate for the minimum frequency deviation that a human could notice.  A more aggressive estimate might be the full-width-at-half-maximum of the prominent measured frequency peaks, which as shown below are closer to 1 Hz, and would lead to bounds three times tighter than the ones presented in this paper.}  We calculated this perturbation size for each chamber wall that the box model predicted was related to a prominent frequency peak in the Hypogeum-wide mean spectrum.  This produces estimates of the distances that individual chamber walls could be perturbed before prominent spectral peaks in the Hypogeum-wide mean spectrum would change by human-noticeable amounts.  In other words, it produces bounds on how much individual chamber wall dimensions could change before the Hypogeum-wide mean spectrum would be noticeably affected.

Finally, to validate the box model predictions, we perturbed several chamber wall dimensions by the amount predicted by the box model, and re-simulated the 3D wave equation in a perturbed version of the full non-rectangular geometry.  Perturbations were generated by translating and then remeshing individual chamber walls in MeshLab \cite{MeshLab}.  The perturbed spectra were compared to the unperturbed spectra to look for the predicted frequency peaks shifting by approximately 3 Hz, as predicted by the box model.

\section{Results}

Figure \ref{figure2a} shows measured and simulated mean Hypogeum-wide frequency spectra.  Agreement between measurement and simulation is reasonable, both qualitatively in overall spectral shape and quantitatively in several prominent frequency peaks, to within the precision of the measurements.  This helps to motivate our use of simulation to probe the effects of perturbing the site’s geometry, which is the only purpose and application for our use of simulation in this paper.  However, we note that our measurements are only able to resolve the amplitudes of prominent frequency peaks to within ~10 dB in most cases, which may mean we are not sensitive to some potential amplitude effects, as we will discuss further below.

Prominent frequency peaks include 37.2, 41.0, 46.1, 50.4, 57.1, 64.3, 72.7, 81.8, and 92.5 Hz.  Previous studies had identified resonances at 70 and 82 Hz \cite{Debertolis, Reznikoff2014}, which are close to the 72.7 and 81.8 Hz peaks we identify, although our resolution is considerably higher.  These peaks represent strong well-defined characteristic frequencies common to multiple non-contiguous chambers in the Hypogeum middle level, and can therefore be understood as acoustic features of the middle level as a whole.

\renewcommand{\thefigure}{2(a)}
\begin{figure}[h!]
\centering
\includegraphics[width=0.5\textwidth]{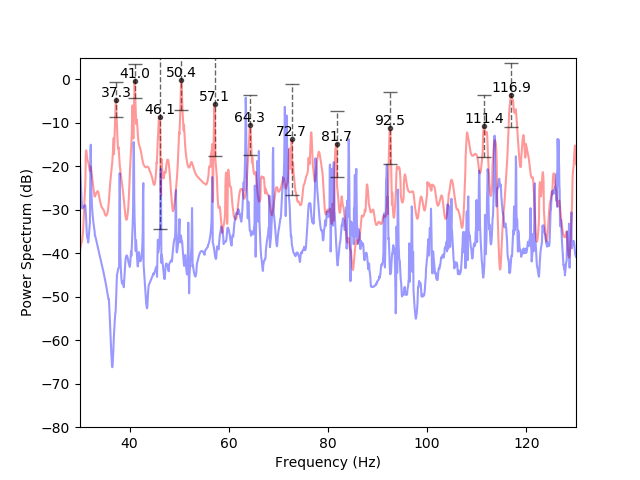}
\caption{Mean measured (red) and simulated (blue) frequency spectrum of the Hypogeum middle level, weighted so that all chambers contribute equally.  Prominent peaks in the measured spectrum are labeled.  Error bars (black) shown for prominent measured frequency peaks reflect measurement-to-measurement spectral variation.}
\label{figure2a}
\end{figure}

Figure \ref{figure2b} shows a measured mean Hypogeum-wide spectrogram, showing that the prominent characteristic frequency peaks are largely time-independent.  Characteristic peaks are present for between 5 and 8 seconds, and were audible during the field work.

\renewcommand{\thefigure}{2(b)}
\begin{figure}[h!]
\centering
\includegraphics[width=0.5\textwidth]{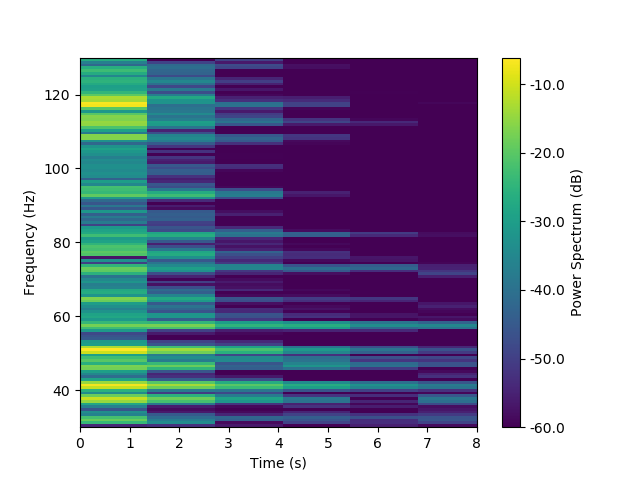}
\caption{Mean measured spectrogram of the Hypogeum middle level.}
\label{figure2b}
\end{figure}

Figure \ref{figure3} shows measured and simulated mean individual chamber frequency spectra for three representative chambers:  27 (a), 18 (b), and 25 (c).  Agreement between measurement and simulation is qualitatively reasonable, suggesting the prominent frequency peaks in the measurements represent room modes, since the simulations incorporate only geometry, with no material properties or absorption.  Box-model modal frequencies are also shown, and agree reasonably with prominent measured and simulated frequency peaks.  The degree of box-model agreement was unexpected given the highly non-rectangular shapes of the chambers.  This motivates our use of the highly-simplified box model to estimate chamber wall perturbation sizes below, which is the only use to which we put the box model in this paper.

Several prominent peaks that appear in measurement and simulation but that do not appear in box-model predictions are room modes of other chambers.  For example, Figures 3(a) and 3(c) both show a prominent peak at 37 Hz that does not appear to be a room mode for either chamber 27 or 25.  This peak is actually a room mode of nearby chamber 20.  Similarly, Figure 3(c) shows a prominent peak at 41 Hz for chamber 25, which can be seen in Figure 3(b) as a room mode of chamber 18, and is also a mode of chamber 20.

\renewcommand{\thefigure}{3}
\begin{figure}[h!]
\centering
\includegraphics[width=0.5\textwidth]{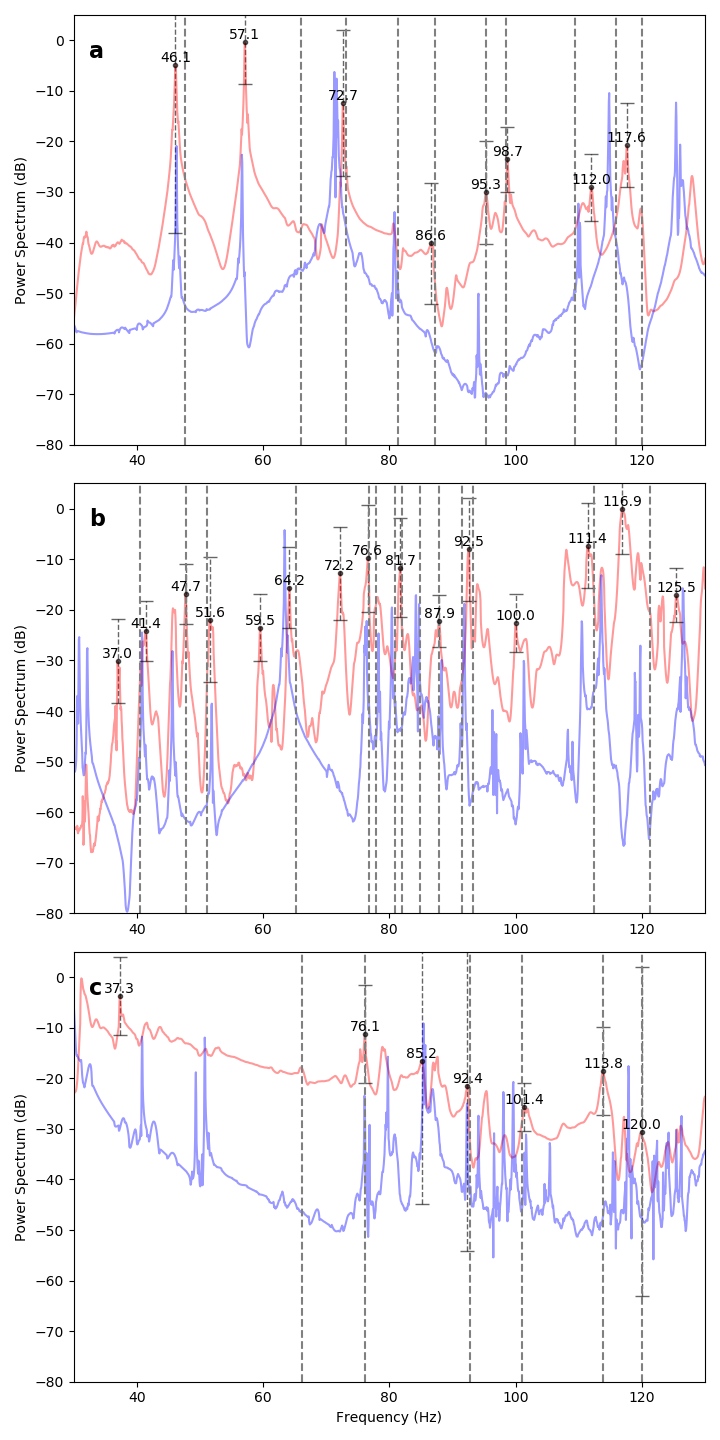}
\caption{Measured (red) and simulated (blue) frequency spectra for representative middle level chambers 27 (a), 18 (b), and 25 (c).  Box-model modal frequencies are shown as dashed grey vertical lines.  Prominent peaks in the measured spectrum are labeled.  Error bars (black) shown for prominent measured frequency peaks reflect measurement-to-measurement spectral variation.}
\label{figure3}
\end{figure}

Table \ref{table1} shows prominent measured frequency peaks from the mean Hypogeum-wide spectrum, with the box-model modes and implied bounds on individual chamber wall dimensions associated with each.  Associated box-model modes are determined by selecting the closest box-model mode within a 3 Hz just-noticeable difference of the measured peak frequency.  The tightest bounds on individual chamber walls are generally in the 10-15 cm range, indicating that those chamber walls could not be moved by more than 10-15 cm before a human would notice a change in the chamber’s frequency spectrum.  The chamber would then sound “out of tune” with the rest of the Hypogeum middle level chambers and mean spectrum.

Each of the prominent measured frequency peaks from the mean Hypogeum-wide spectrum come from box-model room modes from at least two separate chambers, and sometimes as many as five.  Although in cases such as 41.0 Hz this comes from three separate chambers happening to have a single wall of approximately 420 cm in length, more commonly the peaks come from the coincidence of multiple box modes involving multiple walls of differing lengths in multiple non-contiguous chambers.  For example, the 72.7 Hz peak is from the coincidence of three distinct modal arrangements involving four walls in total, the 81.8 Hz peak is from three distinct modal arrangements involving six walls in total, and the 92.5 Hz peak is from five distinct modal arrangements found in five different chambers involving eight walls in total.

\renewcommand{\thetable}{1}
\begin{table}[h!]
\centering
\caption{Prominent measured frequency peaks common to multiple individual chamber spectra, with associated box-model modes and implied bounds on chamber wall dimensions.  Box-model modes are chosen to be within 1 just-noticeable difference (3 Hz) of the measured peak frequency.}
\begin{tabular}{l l l l}
\hline
\parbox[t]{2cm}{Peak \\ Frequency \\ (Hz)} & Chamber & \parbox[t]{2cm}{Box-Model \\ Mode and \\ Modal \\ Frequency (Hz)} & \parbox[t]{2cm}{Implied \\ Bounds on Box-Model \\ Wall \\ Dimensions (cm)} \\ \hline
41.0 & 18 & (0,1,0), 40.4 & 424 $\pm$ 31.4 \\ \cline{2-4}
& 20 & (0,1,0), 40.8 & 420 $\pm$ 30.9 \\ \cline{2-4}
& 24 & (1,0,0), 40.1 & 428 $\pm$ 32.0 \\ \hline
46.1 & 18 & (1,1,0), 47.9 & 670 $\pm$ 146.8 \\
& & & 424 $\pm$ 37.2 \\ \cline{2-4}
& 27 & (1,0,0), 47.6 & 360 $\pm$ 22.7 \\ \hline
50.4 & 18 & (2,0,0), 51.2 & 670 $\pm$ 39.3 \\ \cline{2-4}
& 26 & (1,0,0), 51.2 & 335 $\pm$ 19.6 \\ \hline
57.1 & 20 & (1,1,0), 55.3 & 459 $\pm$ 54.6 \\
& & & 420 $\pm$ 41.8 \\ \cline{2-4}
& 24 & (0,0,1), 56.6 & 303 $\pm$ 16.1 \\ \hline
64.3 & 18 & (2,1,0), 65.2 & 670 $\pm$ 50.0 \\
& & & 424 $\pm$ 50.7 \\ \cline{2-4}
& 25 & (1,0,0), 66.2 & 259 $\pm$ 11.7 \\ \cline{2-4}
& 26 & (0,1,0), 63.5 & 270 $\pm$ 12.8 \\ \cline{2-4}
& 27 & (0,1,0), 66.0 & 260 $\pm$ 11.8 \\ \hline
72.7 & 20 & (2,0,0), 74.7 & 459 $\pm$ 18.4 \\ \cline{2-4}
& 24 & (0,1,1), 70.7 & 405 $\pm$ 47.9 \\
& & & 303 $\pm$ 20.1 \\ \cline{2-4}
& 26 & (0,0,1), 73.0 & 235 $\pm$ 9.7 \\ \cline{2-4}
& 27 & (0,0,1), 73.0 & 235 $\pm$ 9.7 \\ \hline
81.8 & 18 & (0,2,0), 80.9 & 424 $\pm$ 15.7 \\ \cline{2-4}
& 20 & (0,2,0), 81.7 & 420 $\pm$ 15.4 \\ \cline{2-4}
& 24 & (1,1,1), 81.3 & 428 $\pm$ 65.0 \\
& & & 405 $\pm$ 55.1 \\
& & & 303 $\pm$ 23.1 \\ \cline{2-4}
& 26 & (1,1,0), 81.6 & 335 $\pm$ 31.3 \\
& & & 270 $\pm$ 16.4 \\ \cline{2-4}
& 27 & (1,1,0), 81.4 & 360 $\pm$ 38.7 \\
& & & 260 $\pm$ 14.6 \\ \hline
92.5 & 18 & (2,0,1), 93.3 & 670 $\pm$ 71.5 \\
& & & 220 $\pm$ 10.1 \\ \cline{2-4}
& 20 & (0,1,1), 91.0 & 420 $\pm$ 68.7 \\
& & & 211 $\pm$ 8.7 \\ \cline{2-4}
& 24 & (1,2,0), 93.7 & 428 $\pm$ 74.9 \\
& & & 405 $\pm$ 15.9 \\ \cline{2-4}
& 25 & (0,1,0), 92.7 & 185 $\pm$ 6.0  \\ \cline{2-4}
& 27 & (2,0,0), 95.3 & 360 $\pm$ 11.3 \\ \hline
\end{tabular}
\label{table1}
\end{table}

Figure \ref{figure4} shows simulated perturbed and unperturbed frequency spectra for two representative perturbations (or variations) made to two different walls in chamber 27.  In Figure 4(a) one wall dimension was perturbed from approximately 360 to 385 cm, and in Figure 4(b) one wall dimension was perturbed from approximately 235 to 245 cm.  As shown in Table \ref{table1}, the box model would predict a 3 Hz peak shift from 47.6 Hz to 44.6 Hz for Figure 4(a), and from 73.0 Hz to 70.0 Hz for Figure 4(b).  Simulation results agree well with these box-model predictions, and validate the use of the box model for estimating bounds on individual chamber wall dimensions.

\renewcommand{\thefigure}{4}
\begin{figure}[h!]
\centering
\includegraphics[width=0.5\textwidth]{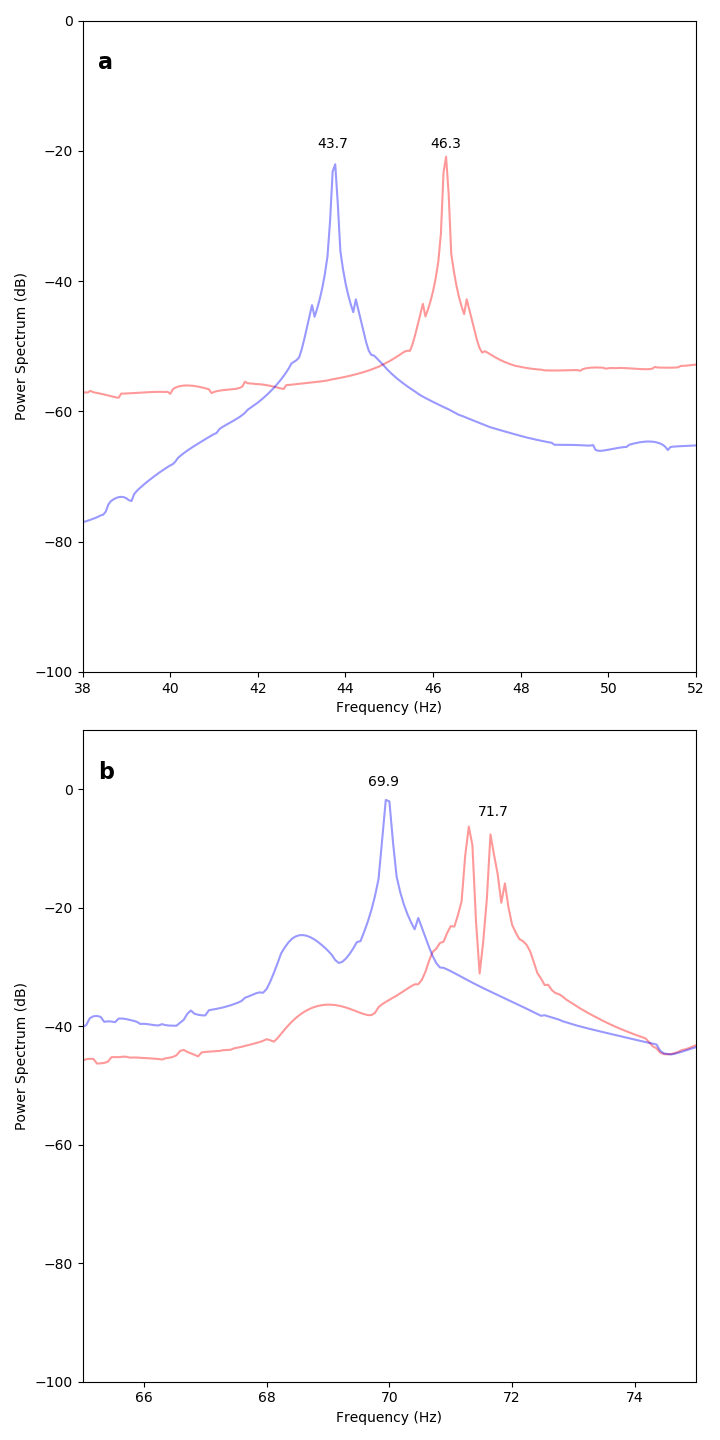}
\caption{Simulated unperturbed (red) and perturbed (blue) frequency spectra for chamber 27.  Perturbations are (a) to move length dimension from 360 to 385 cm, and (b) to move height dimension from 235 to 245 cm.}
\label{figure4}
\end{figure}

\section{Discussion}

There are three pieces of evidence that suggest the frequency spectrum of the Hypogeum was not coincidental.  First, our results in this paper indicate the geometry of the Hypogeum middle level is remarkably finely-tuned.  The aggregate Hypogeum spectrum required jointly fine-tuning the dimensions of multiple distinct walls in multiple non-contiguous chambers to within 10-25 cm, sometimes as many as 6 to 8 walls across 4 to 5 chambers contributing to generate a single aggregate frequency peak.  In other words, there are multiple individual cave walls where if one such wall was moved from its current position by more than about 10-25 cm it would affect the overall frequency spectrum, moving one chamber “out of tune” with the other chambers enough for a human to notice.  This degree of fine-tuning seems unlikely to have come about by coincidence, and suggests that the Hypogeum’s frequency spectrum, or another property closely correlated with the spectrum, strongly influenced how the Neolithic creators shaped the site.

The second piece of evidence is that the frequency spectrum itself looks highly non-random and unusual.  Surprisingly, the ratios of the frequencies of consecutive prominent peaks from 37.2 Hz to 92.5 Hz are nearly all close to the whole-number ratios 9:8 or 10:9, which both characterize the musical “whole tone” interval, as though the Hypogeum’s spectrum defines a musical scale similar to a whole-tone scale.  The Hypogeum therefore appears “tuned” both in the physical and the musical sense.  A whole-tone scale is notably symmetrical and evenly-spaced, not at all what would be expected for a randomly-generated spectrum, and suggests the frequency spectrum held cultural significance.

The third piece of evidence is that the frequency spectrum is a highly unusual and conspicuous feature of a culturally significant site.  The acoustic properties of the Hypogeum have been readily audible even to untrained visitors since the site’s excavation \cite{Mifsud, Evans, Zammit, Griffiths}, and it is reasonable to assume Neolithic people would have been closely attuned to the acoustics of the site.  The process of hand-sculpting the Hypogeum using stone and bone tools and without natural light was likely very loud, and would have excited the acoustics of the site.  Moreover, a frequency spectrum with 30-50 dB peaks would be considered undesirable in a modern room \cite{ISO1996, WHO}.  Indeed, the Hypogeum looks spectrally more similar to an instrument like a bell, gong, or lithophone than to a room or theater \cite{Fletcher}, and the process of shaping the Hypogeum was similar to the process of tuning a bell or gong \cite{Evans, Zammit, Fletcher}.  The Hypogeum was a culturally important site that required significant human effort to carve over many generations and hundreds of years \cite{Trump}.  It seems likely therefore that its unusual conspicuous instrument-like frequency spectrum was a culturally-desired feature, or else it would have been disruptive to any activities that might have taken place.  This requires no assumptions about the kinds of activities that might have occurred in the space, given that during our field work, even minimal human motion like shuffling of feet, sniffing or coughing, or clearing one's throat triggered audible frequency responses from the site.

Taken together, the evidence suggests that spectral properties played some cultural or motivational role in the Hypogeum’s hand-carved geometry.  None of these pieces of evidence necessarily prove that the Neolithic people carved the Hypogeum initially with acoustics in mind.  For example, perhaps the creators came upon some naturally-occurring spectral protofeatures and chose to amplify them (such as the naturally-occurring fissures and bedding planes that outline some walls and floors \cite{Grima}), or perhaps they created or amplified the frequency spectrum while using sound as a heuristic for optimizing some other geometric or spatial features.  Nor can we ascertain the significance of what acoustic properties might have meant to them.  But we do think they preclude coincidence.

One potential objection to this work is that modern acoustic measurements (and models constructed from them) may not be representative of historical conditions.  In particular, modern acoustic measurements are made after the archaeological clearing process undertaken in the early 20th century, and the ancient site may not have been bare rock free of archaeological deposits.  This is a limitation of our methodology.  However, although it is difficult to know for certain given that early excavations of the site were under-documented, there is some reason to believe several of the chambers studied in this work may not have had deposit prior to excavation.  According to the first published documentation on excavations of the Hypogeum by Zammit in 1910, “nearly all the caves, passages, and chambers contained old deposit varying from a few centimeters to over one metre deep” \cite{Zammit}.  However, Zammit also specifies that “the deposit was wanting in the series of caves which are elaborately cut and finished, and in the small caves in the lower storey.”  Most of the chambers studied in this work could reasonably be described as “elaborately cut and finished”, including chambers 18, 20, 24, 26, and 27, which Zammit characterizes using words like “painted”, “elaborate”, “finished”, and “ornamented”.  Among these chambers, Zammit notes that chamber 18 “was found full of old material,” but specifically does not mention anything similar for any of the others.  Furthermore, Zammit notes that several chambers in the lower level were also found comparatively free of deposit.  These chambers are less likely to have been excavated or disturbed during the undocumented excavations prior to Zammit because Zammit states they were found flooded.  Taken together, there is some reason to believe that several of the middle level chambers studied in this work may have been free of deposit prior to excavation, and therefore that modern acoustic measurements might reflect historical conditions.

In this paper we have provided evidence that the peak frequencies of the Hypogeum’s spectrum were tuned by tuning the site’s geometry.  We have focused largely on peak frequencies, and have not evaluated the implications of the observed peak amplitudes beyond noting their magnitude and prominence.  In particular, our measurements are only able to resolve the amplitudes of prominent spectral frequency peaks to a precision of ~10 dB, a level within the bounds of human perception \cite{Kollmeier, ISO1996, WHO}.  It is therefore possible that the Hypogeum’s Neolithic creators might have been attentive to specific amplitude effects that we are unable to resolve in this paper.  Further work is required to increase the measurement precision in order to resolve peak amplitudes to less than ~10 dB, and to investigate any perceptual, acoustical, or archaeological implications of the observed peak amplitudes.

This work identifies the Hypogeum as one of the earliest known examples of a manmade structure with a significant musical element to its interior architecture.  The Hypogeum is one of the earliest surviving manmade structures of any kind, and is unusual among similarly-aged manmade structures because its intact interior allows sonic architectural elements to be studied.  Although there is evidence that prehistoric humans identified, decorated, or culturalized sonically-resonant locations within caves \cite{Rainio, Reznikoff1988, DiazAndreu}, this provides remarkably early evidence of prehistoric humans building such an interior space.

\section{Acknowledgements}

The authors wish to thank Heritage Malta for helpful support and access to the site.  This work was funded by the European Commission through the Marie Skłodowska-Curie Actions program under grant number REP-750618-1.

\section{License}

(c) 2020. This manuscript version is made available under the CC-BY-NC-ND 4.0 license http://creativecommons.org/licenses/by-nc-nd/4.0/

\bibliography{wolfe}

\end{document}